\documentclass{article}

\usepackage{ijcai25}
\usepackage{setspace}
\usepackage{times}
\usepackage{soul}
\usepackage{url}
\usepackage[hidelinks]{hyperref}
\usepackage[utf8]{inputenc}
\usepackage[small]{caption}
\usepackage{graphicx}
\usepackage{amsmath}
\usepackage{amsthm}
\usepackage{booktabs}
\usepackage{algorithm}
\usepackage{algorithmic}
\usepackage[switch]{lineno}
\usepackage{subcaption}
\usepackage{color}

\usepackage{enumitem}

\title{SimTok: Simulating Filter Bubble on Short-video Recommender System with Large Language Model Agents}

\author{
Nicholas Sukiennik$^1$
\and
Haoyu Wang$^1$\and
Zailin Zeng$^{1}$\and
Chen Gao$^2$\and
Yong Li$^1$\\
\affiliations
$^1$Department of Electronic Engineering, Tsinghua University\\
$^2$BNRist, Tsinghua University,\\
\emails
\{sukiennikn10, w-hy23, zengzl23\}@mails.tsinghua.edu.cn,
	\{chgao96, liyong07\}@tsinghua.edu
}

\begin{document}

\maketitle

\begin{abstract} 
An increasing reliance on recommender systems has led to concerns about the creation of filter bubbles on social media, especially on short video platforms like TikTok. However, their formation is still not entirely understood due to the complex dynamics between recommendation algorithms and user feedback. In this paper, we aim to shed light on these dynamics using a large language model-based simulation framework. Our work employs real-world short-video data containing rich video content information and detailed user-agents to realistically simulate the recommendation-feedback cycle. Through large-scale simulations, we demonstrate that LLMs can replicate real-world user-recommender interactions, uncovering key mechanisms driving filter bubble formation. We identify critical factors, such as demographic features and category attraction that exacerbate content homogenization. To mitigate this, we design and test interventions including various cold-start and feedback weighting strategies, showing measurable reductions in filter bubble effects. Our framework enables rapid prototyping of recommendation strategies, offering actionable solutions to enhance content diversity in real-world systems. Furthermore, we analyze how LLM-inherent biases may propagate through recommendations, proposing safeguards to promote equity for vulnerable groups, such as women and low-income populations. By examining the interplay between recommendation and LLM agents, this work advances a deeper understanding of algorithmic bias and provides practical tools to promote inclusive digital spaces.

\end{abstract}

\section{Introduction}\label{sec::intro}

The filter bubble is a phenomenon that has received much attention since the dawn of recommender systems being used filter content on social media platforms, notably being employed as early as 2011 to personalize the Facebook feed \cite{leung2013generational}. The filter bubble is typically defined as the state of being exposed to a narrow scope of content or that which covers only limited set of categories, representing only a small fraction of the possible categories that exist on the platform \cite{pariser2011filter}.
Filter bubbles are concerning due to their implications on both user satisfaction, which has effects on user retention and platform engagement, as well as for their potential to lock users into an echo chamber of information \cite{nguyen_echo_2020}, which could lead to political polarization \cite{lazovich2023filter}. The latter has been cited as a major threat to the normal democratic functioning of society, which is typically premised upon an equitable and free flow of information \cite{Santos_Lelkes_Levin_2021,Vasconcelos_Constantino_Dannenberg_Lumkowsky_Weber_Levin_2021}. The impact that filter bubbles have on users, platforms, and society at large explains why many works have been dedicated to tackling this issue, whether through preventing it or interrupting it during its process of formation. 

The filter bubble has been examined in relation to both traditional social media platforms , long-form video platforms \cite{aridor2020deconstructing} and e-commerce platforms \cite{ge_understanding_2020}. More recently, however, examination of the filter bubble has converged around a new central point, that of short-video platforms, such as TikTok and Kuaishou. In contrast to traditional platforms, which usually have closed friend or following loops that also serve to personalize a user's feed, short video platforms have exploded in popularity due to their ability to recommend desirable content from across all the users of the platform. To make this possible, these platforms have developed expansive use of recommender systems, thereby leading to more potential for the creation of filter bubbles. 
Several works that have addressed the filter bubble on short video platforms have aimed to characterize the phenomenon \cite{piao_humanai_2023,li_exploratory_2022,fu2024heavy}, as well as remediate it through recommender system and algorithm design strategies \cite{li_breaking_2023,li2024fullstage}. As diversity is often viewed as the antithesis to the filter bubble, many works have focused on increasing diversity of recommendations while trying to avoid the trap of the accuracy-diversity dilemma \cite{lu2018diversity,zheng_dgcn_2021,zhou_solving_2010,yang_dgrec_2023,chen_fast_2018}. 

More recently, the rise of large language models (LLMs) has posed new opportunities to gain insights into the workings of social and technical systems, and recommendation is no different. On the social side, recent works have used LLMs to develop simulations of macro-scale political scenarios such as coalition building \cite{moghimifar2024modelling} and diplomacy during wartime situations \cite{hua2024war}, whereas on the micro-scale, they have been used to simulate teamwork scenarios for workplace ideation \cite{shaer2024aiaugmented,he2024ai}, collaboration \cite{guo2024embodied,wang2024unleashing} as well as for primary education \cite{liu2024peergpt}. Meanwhile, they have facilitated simulation of technical systems such as social network behavior \cite{jiang2023socialllm,gao2023s3,wang2024user} and recommender systems \cite{shu2024rah,wang2024user,zhang2023generative}. While works such as \cite{wang2023recagent} and \cite{zhang2023generative}

However, as yet, no works have been dedicated primarily towards the simulation of the filter bubble as an outcome of the interaction between the recommender system and user feedback. In light of the implications of this newfound avenue for behavioral simulation provided by LLMs, this work fills the gap by simulating the user-recommender interface with special focus on the mechanisms that give rise to the filter bubble. LLMs with their complex, often human-like reasoning, can serve as user-agent who can provide realistic feedback in a recommender system scenario, emulating the dynamics of a real system without the need for extensive online testing or real-world datasets. Due to short-video platforms being the foremost media of content recommendation with the rise of TikTok, Kuaishou, Reels, and YouTube shorts, we focus on simulating the formation of the filter bubble in the short-video scenario. 
Specifically, our work makes the following contributions: 
\begin{itemize}[leftmargin=*]
    \item We propose and implement a simulation framework that integrates real-world short-video data, recommendation algorithms, and artificial user-agents to reproduce the dynamics filter bubble as an outcome of the recommender-user feedback interface.
    \item We conduct extensive analysis across dimensions to discover whether and how well the LLM simulation can give rise to realistic filter bubble and how it it influenced by various factors including both user and item characteristics.
    \item We implement a set of recommender system strategies that can serve to alleviate the occurrence of the filter bubble in the simulated scenario, which can, in turn, be used to suggest designs for implementation in real world recommender systems. 
\end{itemize}

\section{Dataset and Methodology}\label{sec:method}

\begin{figure}[t]
  \centering
  \begin{subfigure}{0.22\textwidth}
    \centering
    \includegraphics[width=1\textwidth]{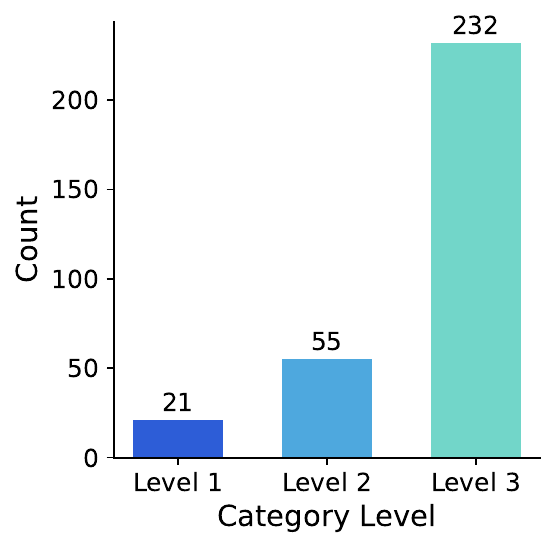}
    \caption{Unique categories.}
    \label{fig:catelevel}
  \end{subfigure}
  \hfill
  \begin{subfigure}{0.22\textwidth}
    \centering
    \includegraphics[width=1.0\textwidth]{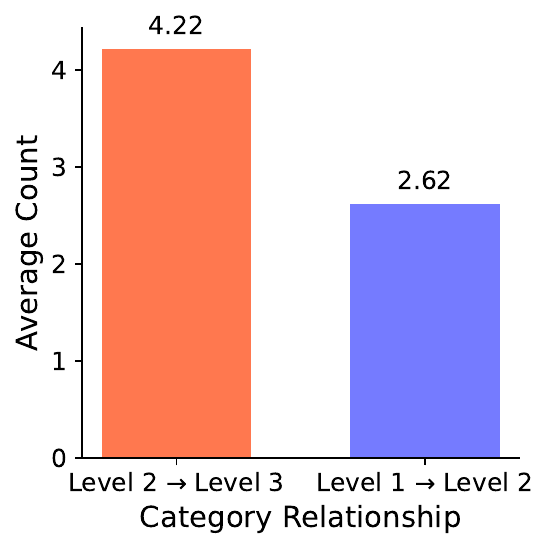}
    \caption{Average child categories.}
    \label{fig:childrenlevel}
  \end{subfigure}
  \caption{Item data statistics.}
  \label{fig:qualifyingdata}
\end{figure}

\begin{figure*}
  \centering
    \begin{subfigure}{0.85\textwidth}
    \centering
\includegraphics[width=\linewidth, trim=0 0 0 0, clip]{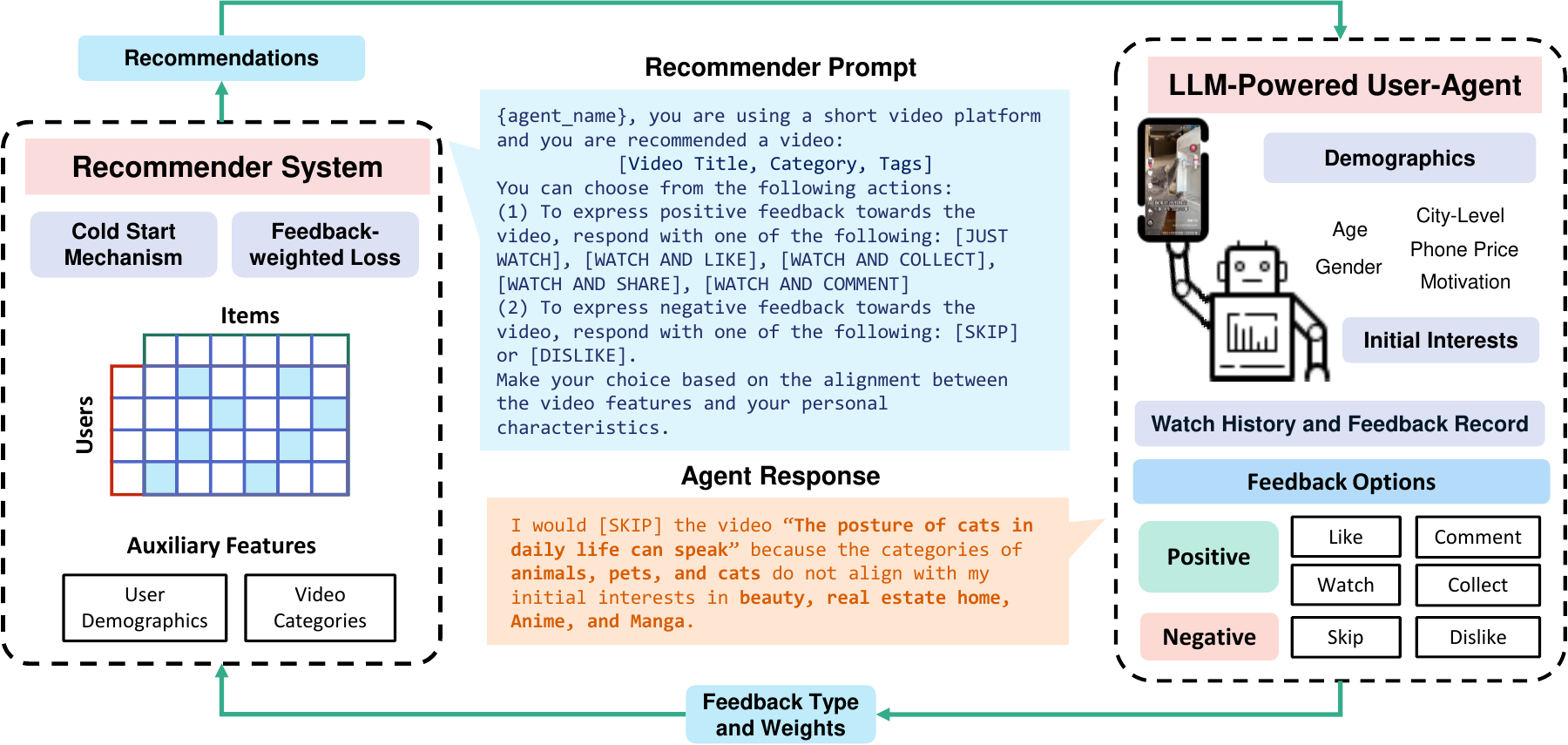} 

  \end{subfigure}
  
  \vspace{-0.2cm}
  \caption{The overview of our simulation framework, which is composed of two major components, the LLM module, which acts as a user-agent, and the recommender module, which generates video suggestions for the user to react to.}

  \label{fig:framework}
\end{figure*}

\subsection{Dataset}

\noindent \textbf{Items.}
The item dataset is adapted from an open-source real-world dataset from one of the top short-video platforms in China. It consists of over 4000 videos with tags, titles, and categories. Each video has three levels of categories, which are arranged in hierarchical structure, where the lower the level, the more fine-grained the category. For example, if the top level category is sports, the second level could be soccer, and the third, Manchester United. As seen in Figure \ref{fig:qualifyingdata}, the items consist of 21 first-level categories, 55 at the second level, and 232 at the third level. Level 1 categories have an average of 2.62 children each whereas Level 2 items have an average of 4.22 children each. 

\noindent \textbf{Users.}
For the user data, we generate artificial user profiles considering the user demographics of age, gender, city-level, phone price, and first-level interest. As a motivator for the agents' actions, we endow them with either a uses and gratifications category or a personality.
The possible uses and gratifications are: Social Interaction, Entertainment, Information-Seeking, Browsing/Variety Seeking, Escapism, which are based on the work of \cite{vaterlaus2021tiktok} who discover the primary gratifications of TikTok users that motivate their platform usage. 

Personality consists of five dimensions: Openness to Experience, Conscientiousness, Extraversion, 
Agreeableness, Neuroticism, adapted from the OCEAN model \cite{egges2003model}. For the purposes of simulation, these dimensions are randomly generated on a scale from 0 to 1 for each user-agent.

\subsection{Simulation Framework}

Figure \ref{fig:framework} shows the overall framework of simulation. The simulation framework seamlessly integrates a large language model module with a recommender module to simulate user-agent interactions. The LLM module generates personalized agent profiles using features such as age, gender, city-level, personality traits, and interests. These profiles are then used by the Recommender module to provide tailored item recommendations, leveraging real-world item data and a collaborative filtering-based recommendation algorithm. 
Agents interact with recommended items through behaviors such as watching, liking, commenting, collecting, skipping, or disliking. Each behavior is assigned a feedback weight, capturing its importance and impact. This feedback is incorporated into the training process of the recommender system, enabling iterative optimization. 
To simulate recommendation our framework leverages two recommendation algorithms: 
\begin{itemize}[leftmargin=*]
    \item \textbf{Matrix Factorization (MF)} \cite{koren2009matrix}: Matrix Factorization leverages the collaborative filtering effect by learning latent embeddings that represent user preferences and item features, enabling the computation of similarities between them. 
    \item \textbf{Factorization Machines (FM)} \cite{rendle2010factorization}: This method is a generalized version of MF which considers additional features of users, items, and context. It serves as one of the most basic models for feature-based recommender systems. Actually, almost all the recent advances could be regarded as the extensions of FM, and thus it is very general and representative. The auxiliary features included in FM training in our framework are: gender, city-level, and phone-price for users, and three categories for each video. 
\end{itemize}

MF is used in our main analysis (Sections \ref{sec:mainresults} and \ref{sec:ablations}) to simulate filter bubble emergence and potential alleviation strategies. Meanwhile, FM is used to incorporate user demographic factors to discover the interplay between the recommender and the user-agent in its potential to propagate bias, as discussed in Section \ref{sec:factors}. FM is more suitable for a bias analysis as it allows the recommender to incorporate both user and item features to model user interest representations in a more intricate way. The use of additional factors, however, introduces opacity in the learned representation, which we aim to disentangle via the analysis in Section \ref{sec:factors}. 

To serve recommendations to the user-agents, our framework employs text summaries of video items, converting them into a prompt, which the user can evaluate based on that video's contents, as well as his or her own profile characteristics, including demographics, motivations, initial interest, and watch history and feedback record.

Regarding the rationality of using text content to substitute the video itself as sufficient to allow users to provide feedback, we include several crucial item details: video title, a tag, and a three-level hierarchy of categories. The video information also includes ``creator popularity" which could be used as an indicator to the user of the strength of a given trend or type of video, thus providing more signals for feedback. 
This is in line with the way many recommendation algorithms themselves recommend content, often not processing the video content itself due to extensive computational demands, but rather typically using substitutes such as the video tags, categories, popularity information, as well as the network effort taken into account via collaborative filtering \cite{aggarwal_content-based_2016}. 

The framework operates as a closed-loop system, where updated agent profiles and feedback continually refine the recommendation process, reflecting the evolving dynamics of user preferences and system adaptation.
\subsubsection{User-agent feedback}
In the design of recommendation systems, user feedback (such as likes, comments, and collections) is often regarded as an explicit or implicit representation of user preferences. However, different types of feedback may carry varying levels of importance in expressing user preferences. In our simulation scenario, the user-agent is presented with a video recommendation and is prompted to provide a form of feedback along with a brief explanation of why that form of feedback was chosen. The feedback form is logged and fed back into the recommender training pipeline, whereas the explanation can be used to derive insights into the agent's reasoning and the factors that influence the decision. The user-agent has the choice of the following types of feedback: 

\begin{itemize}[leftmargin=*]
    \item \textbf{Watch}: Shows interest but only passively.
    \item \textbf{Like}: A more explicit form of interest than watch.
    \item \textbf{Comment}: Typically signifies strong engagement with the content.
    \item \textbf{Collect}: Suggests the content has long-term value to the user.
    \item \textbf{Skip} or \textbf{Dislike}: Indicates dissatisfaction with the content.
\end{itemize}

To better capture these differences in importance during model training, a feedback weighting mechanism is introduced. Feedback weights quantify the significance of user actions into a scalar value (weight) and assign different levels of influence to various feedback types during the loss calculation.

\subsubsection{Incorporation of feedback weights into loss}

In the proposed simulation framework, feedback weights are introduced to account for the varying importance of user behaviors during the training of the recommender model. These weights, denoted as \( w \), are scalars that quantify the significance of each user interaction type, such as ``Like", ``Comment", ``Collect", ``Skip", or ``Dislike". The feedback weight \( w \) is integrated into the loss function to modulate the contribution of each interaction to the optimization process.

The standard Binary Cross-Entropy (BCE) loss function, which evaluates the discrepancy between predicted probabilities and ground truth labels, is modified as follows:

\begin{equation}
   \mathcal{L} = -\frac{1}{N} \sum_{i=1}^{N} w_i \left[ y_i \log(\hat{y}_i) + (1 - y_i) \log(1 - \hat{y}_i) \right], 
\label{eqn:loss}
\end{equation}

\noindent where \( N \) represents the total number of training samples, \( y_i \) denotes the ground truth label for the \( i \)-th sample, \( \hat{y}_i \) is the predicted probability for the \( i \)-th sample, and \( w_i \) refers to the feedback weight associated with the \( i \)-th sample, which reflects the type of user interaction.

By multiplying the BCE loss of each sample with its corresponding feedback weight \( w_i \), the model emphasizes interactions with higher significance (e.g., "Comment" or "Collect") while de-emphasizing less impactful behaviors (e.g., "Like"). Negative feedback such as "Skip" or "Dislike" is represented with negative weights, which inversely influence the loss, guiding the model to penalize recommendations leading to such interactions.
\subsection{Filter Bubble Evaluation Method}
The diagram in Figure \ref{fig:framework} illustrates the explicit working mechanism of our system. Based on this, we evaluate the output through the below methods.
We employ the following methods to evaluate the presence and impact of filter bubbles.
First, we define the following metrics:
\begin{itemize}[leftmargin=*]

    \item \textbf{Overall Coverage} $C_{i,u,l}$: The ratio of the number of unique categories watched by user $u$ at level $l$ in iteration $i$ to the total number of categories among all videos at that level.
    \item \textbf{Overall Entropy} $E_{i,u,l}$: The Shannon entropy of the categories of watched videos at level $l$ for user $u$ in iteration $i$.
    \item \textbf{Satisfaction} $s_{i,u}$: For iteration $i$ and user $u$, it is the ratio of the number of positively responded videos to the total number of videos watched at a given iteration.

\end{itemize}

Namely, the three indicators defined to quantify the state of diversity and the filter bubble throughout the simulation process are defined as follows.

Coverage is calculated as the number of categories seen by a user out of all the possible categories at a given level:
\begin{equation}\label{eqn:naivecoverage1}
\mathcal{C}_l = \frac{n_{\text{seen},l}}{n_{\text{total},l}},
\end{equation}
where $\mathcal{C}_l$ represents the coverage at a specific level, $n_{\text{seen},l}$ denotes the number of categories observed at that level, and $n_{\text{total},l}$ refers to the total number of categories available at that level.

Entropy is defined as:
\begin{equation}
E_{i,u,l} = - \sum_{c \in \mathcal{C}_{i,u,l}} p(c) \log p(c),
\end{equation} where
$\mathcal{C}_{i,u,l}$ represents the set of unique video categories watched by user $u$ at level $l$ in iteration $i$, $
p(c)$ is the probability of a video belonging to category 
$c$, calculated as the frequency of category c divided by the total number of videos watched by user $u$ at level $l$ in iteration $i$. Entropy in this scenario is used as a measure of the uncertainty of categories present in a given list of watched videos, where the higher the entropy, the more uncertain. Therefore, it is a good measure of the diversity of a user's exposed video list. 

Satisfaction, in turn, is quantified as 
\begin{equation}
s_{i,u} = \frac{n_{\text{+},i,u}}{n_{\text{total},i,u}},
\end{equation}

\noindent where $s_{i,u}$ is satisfaction for iteration $i$ and user $u$, $n_{\text{+}}$ is the number of videos with positive responses by user $u$ in iteration $i$, ${\text{total},i,u}$ is the total number of videos watched by user $u$ in iteration $i$.

To determine whether a user is classified as being impacted by a filter bubble during each iteration, we define a coverage-based criterion. Specifically, we set a threshold: if the number of categories accessed by a user falls below the median number of categories accessed by all users at a given level, the user is classified as being "in" the filter bubble at that level. Mathematically, this is expressed as:
\begin{equation}\label{eqn:criterion}
F_{u,t} =
\begin{cases}
\text{"in"}, & \text{if } A_{u,t} < \text{Median}(A_{\cdot,t}) \\
\text{"out"}, & \text{otherwise}
\end{cases},
\end{equation}
where $F_{u,t}$ indicates the filter bubble status of user $u$ at time window $t$, identifying whether the user is being impacted by the filter bubble. Here, $A_{u,t}$ represents the number of distinct categories encountered by the user during time window $t$, and $\text{Median}(A_{\cdot,t})$ denotes the median number of categories encountered by all users in the same time window. Using this classification, we compute the proportion of users classified as being impacted by the filter bubble for each level at each iteration throughout  the simulation period.

\section{Experiments and Results}\label{sec::experiments}

\subsection{Reproducing the Filter Bubble}\label{sec:mainresults}

The first task of our study is to reproduce the filter bubble using our simulation framework. Below, we present the results of simulation experiments with the two user motivation types separately. The metrics for evaluating the extent of the filter bubble are entropy and coverage. The number of user agents is 20 and each one is recommended 5 items per iteration, in series, which they can choose to offer positive or negative feedback, according to the paradigm seen in Figure \ref{fig:framework}.

\subsection{Uses and Gratifications}

\begin{figure}[t!]
  \centering
  \begin{subfigure}{0.23\textwidth}
    \centering
    \includegraphics[width=1\textwidth]{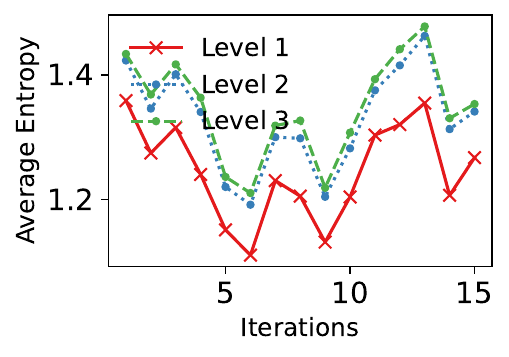}
        \vspace{-0.3cm}
    \caption{Entropy over time per level.}
  \end{subfigure}
  \hfill
  \begin{subfigure}{0.23\textwidth}
    \centering
    \includegraphics[width=1\textwidth]{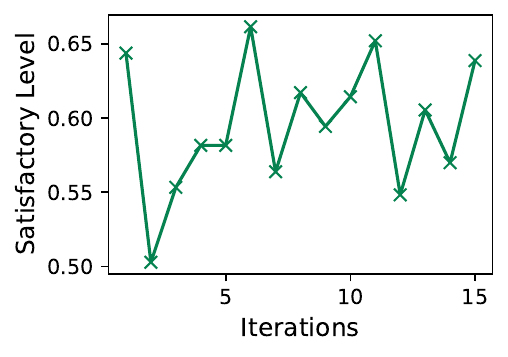}
        \vspace{-0.3cm}
    \caption{Satisfaction level over time.}

  \end{subfigure}
    \vspace{-0.3cm}
  \caption{Coverage and satisfaction over time for users motivated by uses and gratifications.}
  \vspace{-.2cm}
  \label{fig:traditionalfb_m}
\end{figure}

In Figure \ref{fig:traditionalfb_m}, the trend of entropy as it progresses throughout the simulation is displayed, as well as the user satisfaction, which is measured by the average number of instances of positive feedback per epoch over all users. We can see that entropy drastically decreases in the first few epochs, then gradually recovers to a degree. At the end filter bubble is most severe at level 1, and less severe at levels 2 and 3. We note that satisfaction basically follows the trend of diversity at all times, where the less diverse, the less satisfied the users are. This is reasonable considering that too much exposure to a specific type of content can lead to boredom, as addressed in \cite{li2023breaking}.

\begin{figure}[t!]
  \centering
    \centering
    \includegraphics[width=0.9\linewidth]{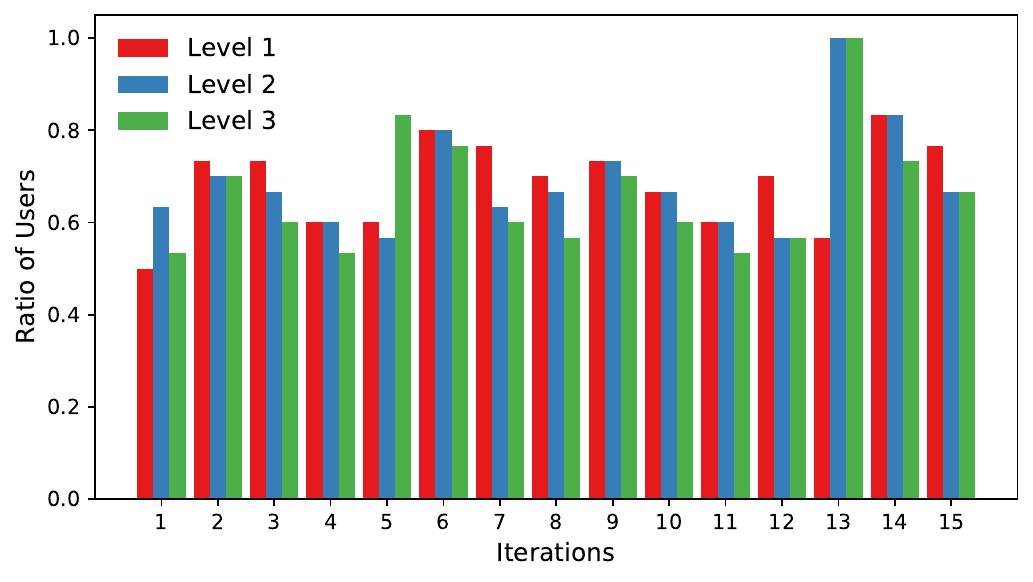}

  \vspace{-0.5cm}
  \caption{Filter bubble changes over time for three category levels for users motivated by uses and gratifications.}
  \label{fig:feedback_bubble_m}
\end{figure}

Figure \ref{fig:feedback_bubble_m} shows the evolution of filter bubble over time, as per the criterion in Equation \ref{eqn:criterion}. We can see that the nature of the filter bubble changes over time, with it starting low for all levels, then slowly increasing. Whereas initially level 2 has the most severe filter bubble, at the end the filter bubble at level 1 is most severe, and levels 2 and 3 also increase in severity.

\subsection{Study on Personality}
\begin{figure}[t!]
  \centering
  \begin{subfigure}{0.23\textwidth}
    \centering
    \includegraphics[width=1\textwidth]{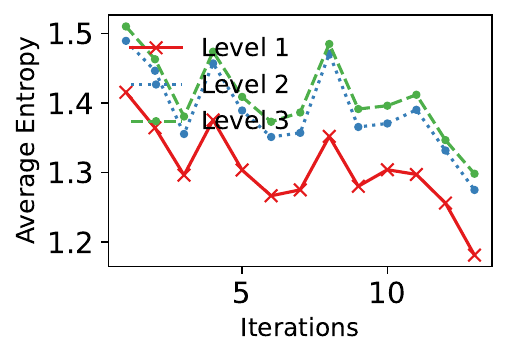}
        \vspace{-0.3cm}
    \caption{Entropy over time per level.}
  \end{subfigure}
  \hfill
  \begin{subfigure}{0.23\textwidth}
    \centering
    \includegraphics[width=1\textwidth]{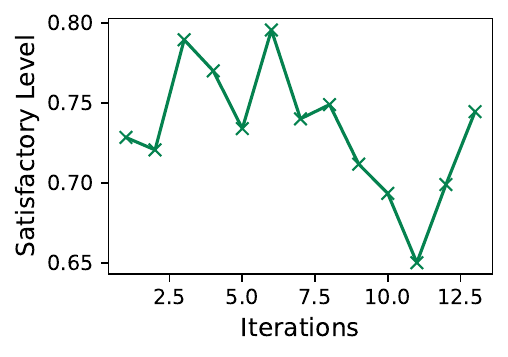}
         \vspace{-0.3cm}
    \caption{Satisfaction level over time.}

  \end{subfigure}
     \vspace{-0.3cm}
  \caption{Coverage and satisfaction over time for users motivated by personality.}
  \vspace{-.2cm}
  \label{fig:traditionalfb_p}
\end{figure}

In this section, we display the diversity and filter bubble results with regard to users motivated by personality, 
In Figure \ref{fig:traditionalfb_p}, the trend of entropy as it progresses throughout the simulation is displayed, as well as the user satisfaction. We can see that entropy drastically decreases in the first few epochs and remains low throughout. This tells us that personality as a motivator serves to make the LLM-based agents act in a way that is more conducive to filter bubble formation. Satisfaction similarly drops over time but makes a rebound at the last iteration, which may be an aberration. 

\begin{figure}[t!]
  \centering
    \centering
    \includegraphics[width=0.9\linewidth]{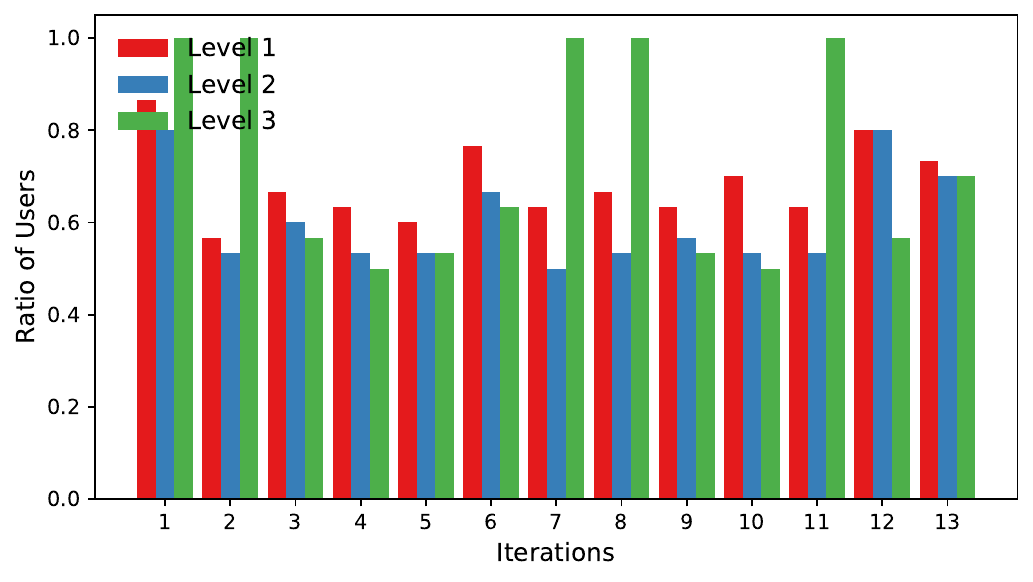}

    \vspace{-0.5cm}
  \caption{Filter bubble changes over time for three category levels for users motivated by personality.}
  \label{fig:feedback_bubble_p}
\end{figure}

Figure \ref{fig:feedback_bubble_p} shows the evolution of filter bubble over time, as per the criterion in Equation \ref{eqn:criterion}. We can see that the nature of the filter bubble changes over time, with it starting high for all levels, dropped in the middle, and then increasing towards the end. Whereas initially level 2 has the most severe filter bubble, at the end the filter bubble at level 1 is most severe, and levels 2 and 3 also increase in severity.

\subsection{Factors Influencing Filter Bubble Formation}\label{sec:factors}

\begin{figure*}
    \centering
    \begin{subfigure}[b]{0.245\textwidth} %
        \includegraphics[width=\textwidth]{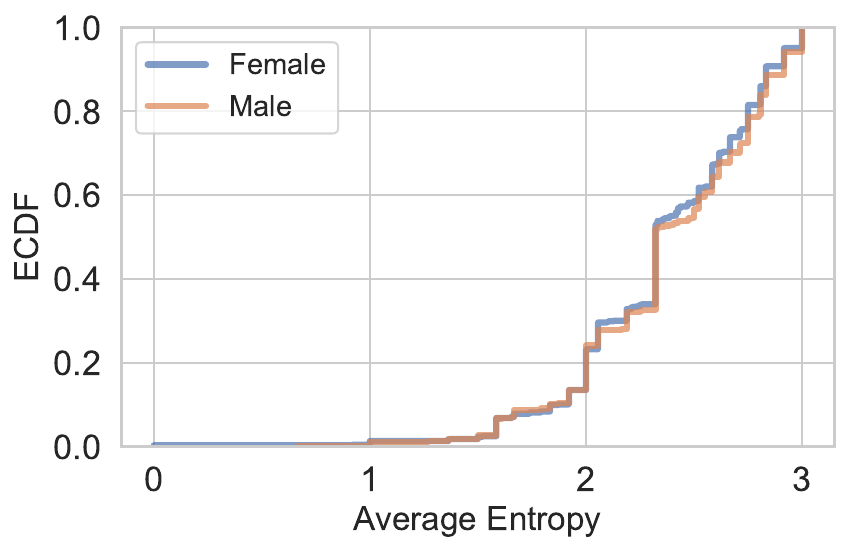}
    \end{subfigure}
    \hspace{-0.15em} %
    \begin{subfigure}[b]{0.245\textwidth}
        \includegraphics[width=\textwidth]{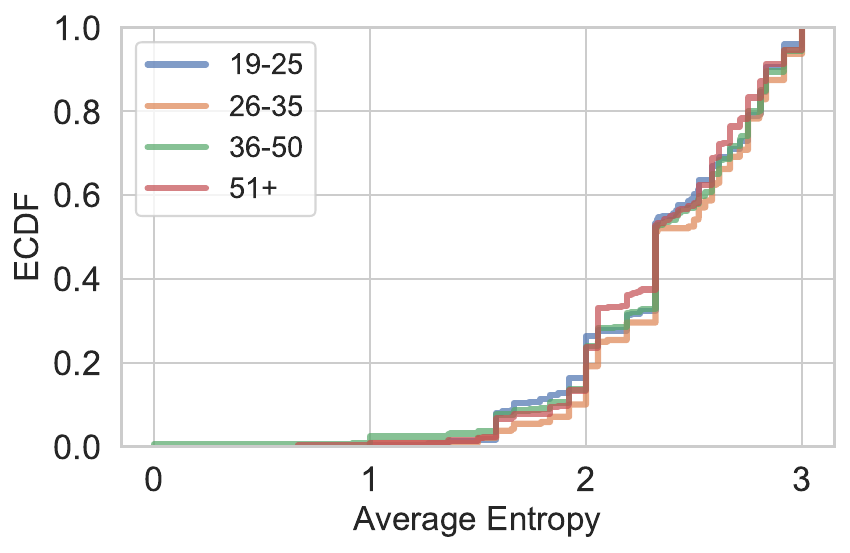}
    \end{subfigure}
    \hspace{-0.15em}
    \begin{subfigure}[b]{0.245\textwidth}
        \includegraphics[width=\textwidth]{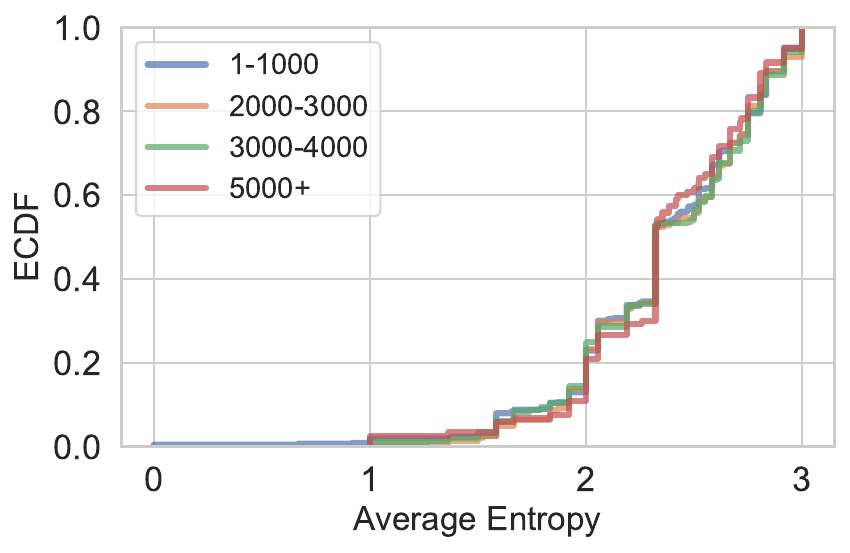}
    \end{subfigure}
    \hspace{-0.15em}
    \begin{subfigure}[b]{0.245\textwidth}
        \includegraphics[width=\textwidth]{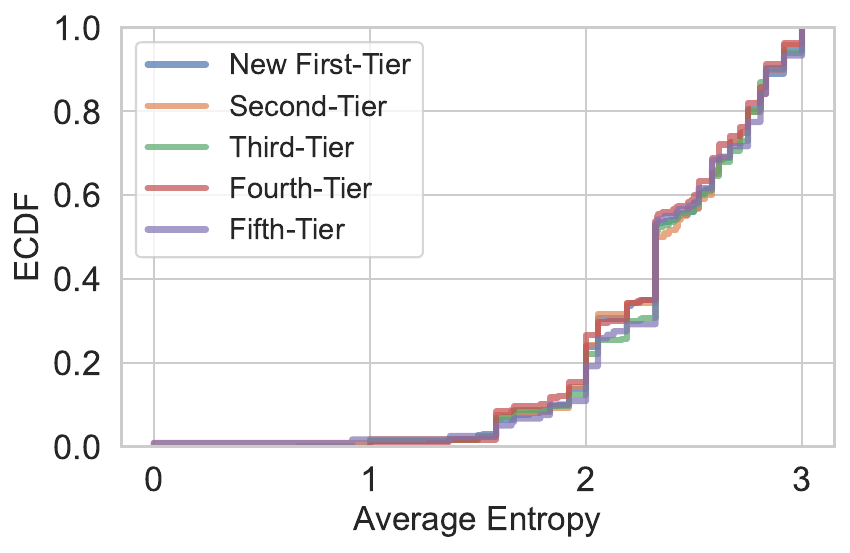}
    \end{subfigure}
    \caption{Coverage distributions for different user demographic features.}
    \label{fig::demog}
\end{figure*}

\noindent \textbf{User Factors.}
Figure \ref{fig::demog} shows the influence of the different agent demographic features on diversity, namely age, phone price, gender, and city level, where city-level corresponds to a level of economic development where the higher, the more developed. Phone price, on the other hand, is an indicator of a user's income, and is a realistic proxy given that this data can really be obtained via platforms, whereas real user income data can be very difficult to collect. The figure shows and Empirical Cumulative Probability Function (ECDF) over entropy for each demographic feature to determine which factors have the biggest effect on filter bubble formation. Although the differences are small, we do note that there are some notable disparities in entropy for the various demographic groups. Namely, the most drastic influence category levels comes from age and phone price. One example of phone price having an impact on user content preferences can be seen in a specific example in user feedback, wherein a user said, "I would skip this video because even though it is about advanced digital products, the title suggests that it is about a cheap and poor-quality phone, which is not aligned with my willingness to spend RMB 2000-3000 on a phone." In this way, phone price can be interpreted to be a meaningful feature guiding content interests. With respect to age, older user-agents have a tendency towards higher coverage whereas younger user-agents tend towards lower diversity. Although gender does not have a large impact, it still shows slightly increased diversity for females at all levels. Finally, users in first-tier cities have higher coverage in category levels 1 and 2, but in level 3, users in the first-tier have somewhat lower diversity than lower tiers such as third. These correlations show us that the simulator picks up on demographic implications in their influence towards user preferences and the likelihood of filter bubble for specific user groups, showing us that LLM agents have nuanced characteristics that can give rise to personalized behaviors when prompted in a detailed fashion.

\section{Alleviating the Filter Bubble}\label{sec:ablations}
In this section, we conduct several ablation studies on different mechanisms of the simulator with the aim of discovering which components can help alleviate the formation of the filter bubble. 
\subsection{Cold Start Matching Strategy}

The cold start category matching ratio is intended to represent the feature on short-video platforms that asks users to specify their interests explicitly. We represent this in the simulation model by randomly assigning each user-agent with three initial interests that come from the existing top level categories among the videos in the item data. These specified interests are then used in the user profile prompts by telling the user their "initial interested categories". They are specified as "initial" to allow for the possibility of changing interests over time. More specifically, the cold start category matching ratio describes the proportion of videos that align with users interests in the first iteration, as seen in $\text{CSCMR} = \frac{V_{\text{aligned}}}{V_{\text{total}}}$,
where CSCMR is cold start category matching ratio, $V_{\text{aligned}}$ represents the number of videos in the first iteration that align with users' interests, and $V_{\text{total}}$ represents the total number of videos shown to users in the first iteration. The default cold start ratio is 50\%, and we also test 0\% (no cold start mechanism) 25\%, 75\%, and 100\% (where all the items in the first iteration are aligned with users' interests). 

\begin{figure}[t!]
  \centering
  \begin{subfigure}{0.23\textwidth}
    \centering
    \includegraphics[width=1\textwidth]{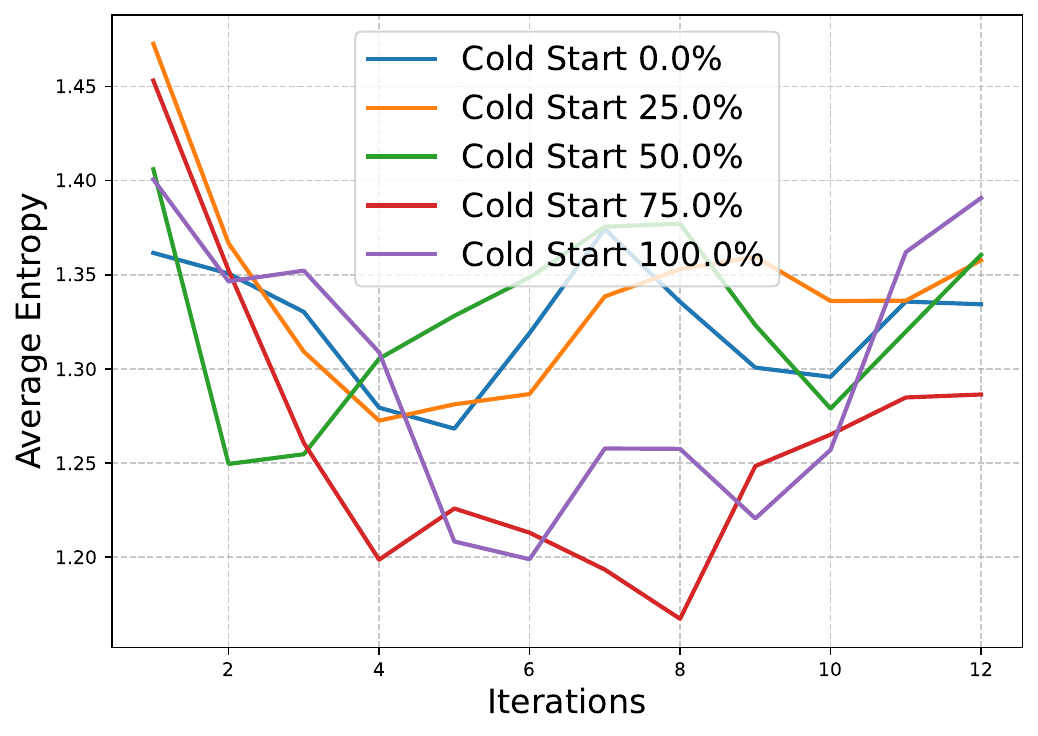}
    \caption{Entropy over time.}
  \end{subfigure}
  \hfill
  \begin{subfigure}{0.23\textwidth}
    \centering
    \includegraphics[width=1\textwidth]{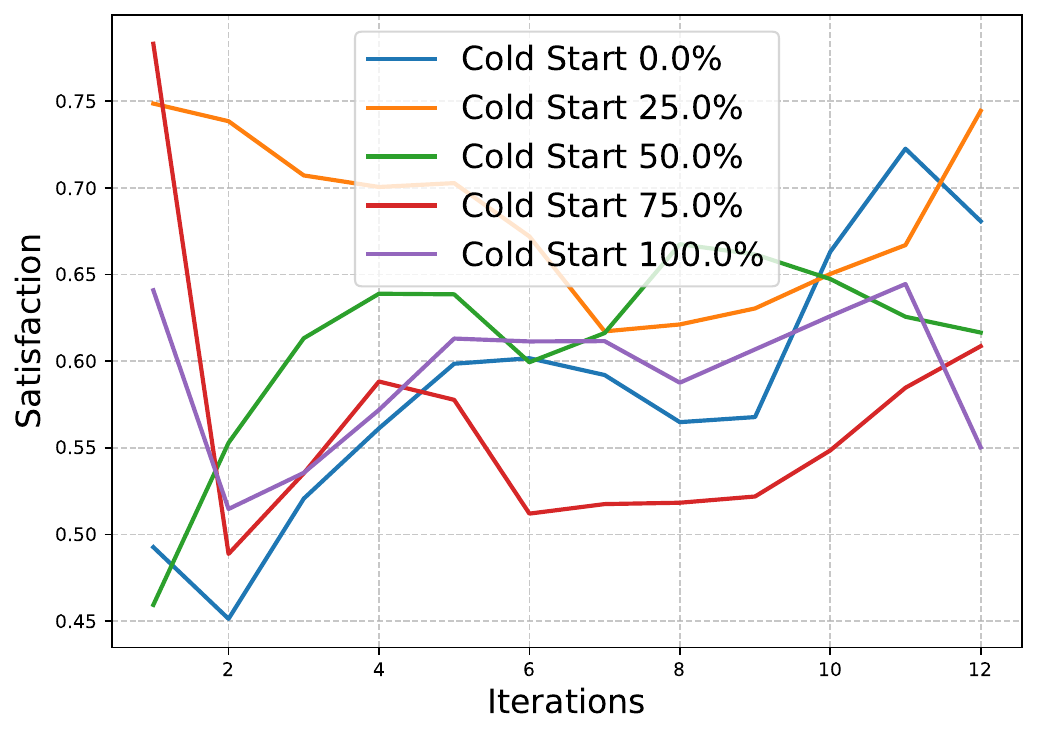}
    \caption{Satisfaction level over time.}

  \end{subfigure}
  \caption{Comparing the cold start category matching ratio and its impact on diversity (entropy) and user satisfaction over time. }
  \vspace{-.2cm}
  \label{fig:coldstart}
\end{figure}

We note from Figure \ref{fig:coldstart} that a CSCMR of 100\% leads to the highest diversity and therefore lowest filter bubble occurrence by then end of the simulation. However, this comes at a cost of satisfaction. On the other hand, a CSCMR of both 0\% (no cold start) and 25\% increase entropy substantially while also maintaining high satisfaction, although satisfaction is consistently higher at the 25\% value. 

These findings show that users' initial, explicitly specified interests on a platform can be utilized in different ways, and depending on how they are used, the progression of filter development can be affected. The reason a lower cold start matching ratio leads to more diversity and higher satisfaction, we believe, is because it allows for the possibility of serendipity, or viewing videos that are in a liked category that was not known about before \cite{kaminskas2016diversity}. This allows plenty of room for user interests to expand and evolve based on exposed videos that are in categories that are not contained in the user's list of initial interests. Through the process of simulation, we indeed note several intriguing instances of interest evolution that match what users may experience on a real platform. One example is a user-agent who has an initial interest of ``history", but after watching one video about Soviet leaders, he decided to skip the next video about history because it was not about Soviet leaders, an interest he had developed through exposure to more fine-grained categories within existing interests. 

\subsection{Feedback Weighting Strategy}

\begin{figure}[t!]
  \centering
  \begin{subfigure}{0.23\textwidth}
    \centering
    \includegraphics[width=1\textwidth]{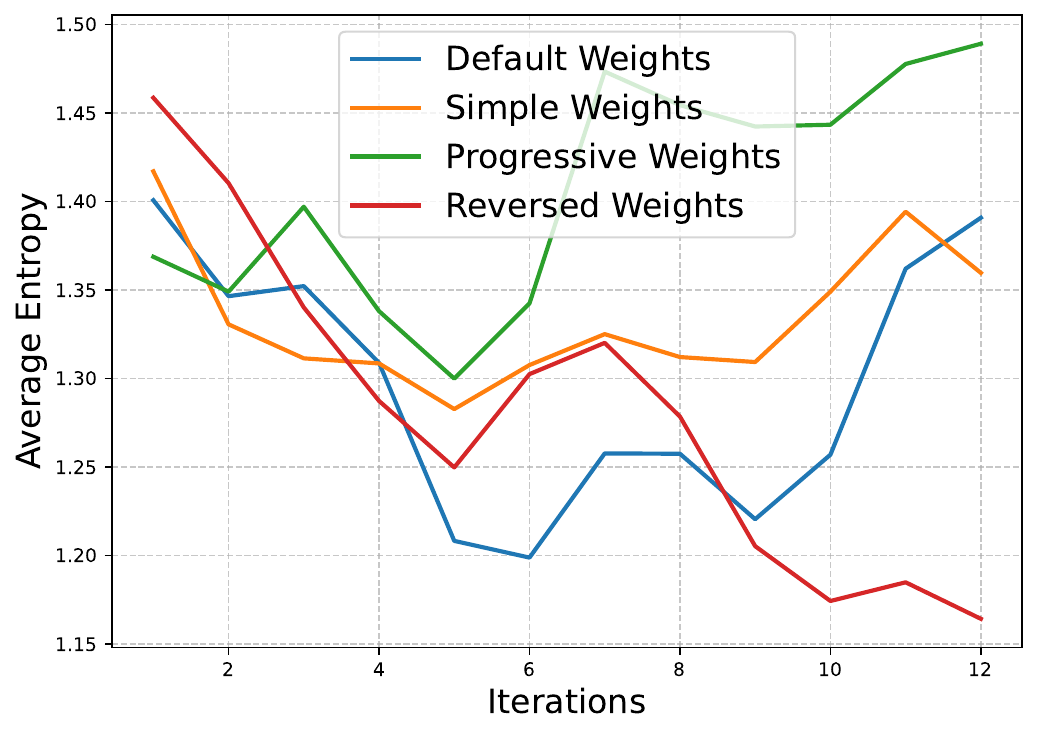}
    \caption{Entropy over time.}
  \end{subfigure}
  \hfill
  \begin{subfigure}{0.23\textwidth}
    \centering
    \includegraphics[width=1\textwidth]{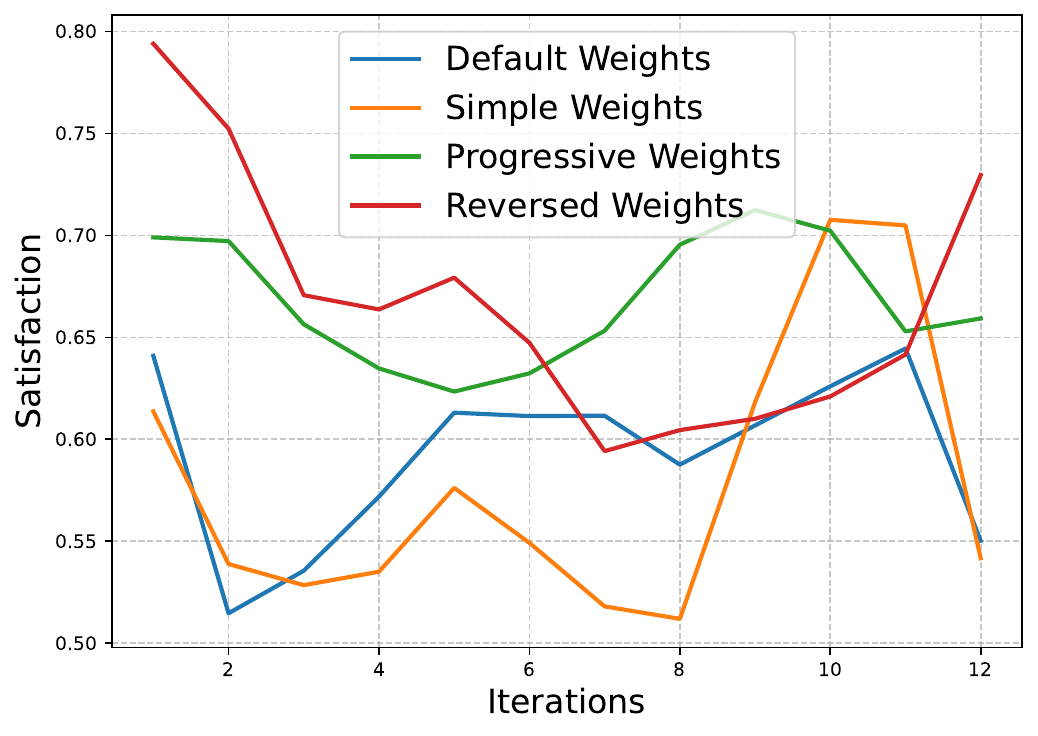}
    \caption{Satisfaction level over time.}
  \end{subfigure}
  \caption{Comparing different feedback weighting strategies and their impact on diversity (entropy) and user satisfaction over time.}
  \vspace{-.2cm}
  \label{fig:weights}
\end{figure}

We also introduce a handful of different feedback weighting strategies where different forms of user feedback are propagated through model training to change the the way item representations are learned when used in conjunction with the loss function in Equation \ref{eqn:loss}, thereby having an impact on downstream recommendations. 
The feedback weights according to each strategy are outlined as follows: 
\begin{itemize}[leftmargin=*]
    \item \textbf{Default Weights}: Positive (JUST WATCH: 1, WATCH AND LIKE: 2, WATCH AND COMMENT: 2, WATCH AND COLLECT: 2), Negative (SKIP: 0, DISLIKE: -1)
\item \textbf{Simple Weights}: Positive (JUST WATCH: 1), Negative (SKIP: 0)
\item \textbf{Progressive Weights}: Positive (JUST WATCH: 1, WATCH AND LIKE: 2, WATCH AND COMMENT: 3, WATCH AND COLLECT: 4), Negative (SKIP: -1, DISLIKE: -2)
\item \textbf{Reversed Weights}: Positive (JUST WATCH: 2, WATCH AND LIKE: 1, WATCH AND COMMENT: 1, WATCH AND COLLECT: 1), Negative (SKIP: 0, DISLIKE: -1)

\end{itemize}
We note from Figure \ref{fig:weights} that when weights are progressive, entropy increases drastically while keeping satisfaction high, proving this to be an effective strategy to alleviate the filter bubble. In comparison, the default weights have lower entropy and also lower satisfaction throughout. Simple weights, which only consider implicit feedback, i.e. watch or skip, results in moderate entropy throughout, and very low satisfaction for most iterations, showing that the inclusion of both implicit and explicit feedback are necessary to accurately model user-agents interests in our simulation, and implicit feedback cannot be relied upon alone. Meanwhile, reversed weights have the highest satisfaction level with lowest entropy, showing that they are not an effective strategy for filter bubble alleviation.

\vspace{-0.5em}

\section{Related Work}\label{sec::related}
The related works for this paper can be divided into two categories: filter bubble and agent-based simulation.

\subsection{Filter Bubble}
With the development of online services such as online social networks, online shopping, and short videos, the amount of information in cyberspace has been growing larger and larger, exceeding the range that ordinary users can handle. Against this backdrop, service providers deploy personalized recommendation algorithms in their systems to filter information. Specifically, the recommendation algorithms filter out content that users may be interested in from vast amounts of information based on users' historical behaviors, profiles, and other information~\cite{wu2022survey}. After users interact with the output of recommendation algorithms, more new behavioral data will be collected and will be further used by the recommendation algorithms to update model parameters, and then update the recommendation results. This process forms a feedback loop~\cite{mansoury2020feedback}, which leads to the recommendation results becoming more and more concentrated and eventually gives rise to the filter bubble~\cite{pariser2011filter}. The existing research on filter bubbles mainly focuses on simple data-based analysis~\cite{philips2024exploring} or attempts to address it from the perspective of diverse recommendations~\cite{zhang2024practical}. In other words, these studies are based on already biased data (the bias here is because the data are always influenced by the already deployed recommendation algorithms).

In this work, our motivation for using LLMs is related to the first type of research. Since large language models can accurately identify current interest needs, the agents constructed with large language models can effectively serve as simulation objects to interact with already deployed recommendation algorithms and generate corresponding user behaviors.

\subsection{Agent-Based Modeling and Simulation}
Agent-based modeling and simulation is a fundamental scientific research method, which is widely used in fields such as complex systems, social networks, and user behavior analysis~\cite{helbing2012agent}. Generally speaking, agent-based modeling and simulation drives the behaviors of agents by defining rules or building models, and further observes different types of behaviors and patterns at the macro level.
Although agent-based modeling and simulation is a research field with a long history, the traditional methods still face the key challenge of insufficiently accurate modeling for each agent.
In recent years, the simulation capabilities of large language model agents have led a technological revolution in the field of agent-based modeling and simulation~\cite{gao2024large}. Researchers have applied agents to the simulation of social behaviors~\cite{park2023generative}, economic behaviors~\cite{li2024econagent}, etc. Among them, some studies~\cite{zhang2023generative,wang2023recagent} have already considered using large language model agents for the simulation of user behaviors in recommendation systems. However, such studies lack a profound understanding and consideration of the filter bubble.

Starting from the important issue of filter bubble as social good, this work fully studies the simulation of filter bubble in recommendation systems by large language model agents and reveals the relevant mechanisms.

\section{Discussion, Conclusions and Future Work}\label{sec::conclusion}

Our study shows that LLM agents can be effectively used to simulate the users of a short video platform, exhibiting realistic behaviors with sound reasoning. Our system successfully reproduces the emergence of filter bubbles via simulating the interface between the recommender system and user feedback. The results demonstrate that user diversity, measured by entropy and coverage, can be influenced by two forms of user motivation (uses and gratifications or personality) as well as both user and item features. Specifically, user-agents motivated by personality tend to experience more severe filter bubble effects, with entropy remaining consistently lower compared to those motivated by uses and gratifications. Satisfaction trends align closely with diversity, showing that reduced diversity leads to diminished user satisfaction. Additionally, demographic factors such as phone price show a significant impact on filter bubble formation, while gender has a comparatively minor effect. Moreover, we propose two forms of filter bubble alleviation using cold start and feedback weighting strategies. We find that lower cold start matching ratios lead to higher diversity over time, whereas progressive feedback can also reduce the filter bubble effect. 
This work also has some limitations. Namely, agents are not exposed to the actual video but rather textual data representing the video. As for future work, we could integrate video or image data into a multi-modal LLM pipeline for more accurate feedback on the video content. We could also conduct real-world testing of the interventions in live systems to validate their practical effectiveness.

\newpage
\setstretch{0}
\bibliographystyle{named}
\bibliography{ijcai25}

\end{document}